\documentclass[prl,twocolumn,superscriptaddress,showpacs]{revtex4}

\usepackage{graphicx}
\usepackage{amsmath}

\begin{document}

\newcommand{\Froude}{{\rm Fr}}

\newcommand{\LANL}{Condensed Matter and Thermal Physics $\&$ Center for Nonlinear Studies, Los Alamos National Lab, NM, 87545, USA}

\newcommand{\SZFKI}{Research Institute for Solid State Physics and Optics,
                    POB 49, H-1525 Budapest, Hungary}

\newcommand{\EXXON}{ExxonMobil Research and Engineering, 1545 Route 22 East, Annandale, NJ 08801, USA}

\title{Two scenarios for avalanche dynamics in inclined granular layers}

\author{Tam\'as B\"orzs\"onyi}
\email{btamas@szfki.hu}
  \affiliation{\LANL}
  \affiliation{\SZFKI}

\author{Thomas C. Halsey}
  \affiliation{\EXXON}

\author{Robert E. Ecke}
  \affiliation{\LANL}

\date{\today}

\begin{abstract}
We report experimental measurements of avalanche behavior of thin granular
layers on an inclined plane for low volume flow rate.
The dynamical properties of avalanches were quantitatively and
qualitatively different for smooth glass beads compared to irregular
granular materials such as sand. Two scenarios for granular avalanches on
an incline are identified and a theoretical explanation for these different
scenarios is developed based on a depth-averaged approach that
takes into account the differing rheologies of the granular materials.
\end{abstract}


\maketitle

Avalanche behavior of granular material has many natural realizations from snow avalanches
to massive rockslides.   Physics-oriented avalanche investigations 
focusing on sand-pile kinematics \cite{jali1989} have given way to more recent work, which 
has probed granular interactions and the appropriate balance between continuum and discrete approaches \cite{go2003}.  
Often avalanche dynamics have been studied with a bulk granular substrate,
either in a rotating drum \cite{ra1990,cogo2003,ev1991,jali1989,ra2002a,tevi2002}    
where the critical angle $\theta_c$, {\it i.e.}, the angle where grains start to flow, is reached by the  
slow rotation of the drum, or on a pile \cite{jali1989,ledu2000}.    
Alternatively, one can investigate flow on an inclined plane where an underlying solid surface
constrains the flow.  In inclined layer granular flow at high volume flow rates a
uniformly thick layer forms, whereas for lower flow rates waves in the form of thickness
variations appear \cite{da2001,fopo2003,sa1979,da1990,prpa2000,loke2001}.  
For both the freely flowing and wave modulated cases,
continuum descriptions of the flows based on flow rheology describe the experiments well
\cite{ra2002b,doan1999,emcl2000}. 

In this Letter we focus not on steady or modulated flows but rather on flows exhibiting distinct, well-resolved avalanches. We have explored these flows on an inclined plane and have
discovered two distinct scenarios for the dynamic avalanche behavior, depending on the
character of the grains. For rough non-spherical grains (RNSG), avalanches are faster, 
bigger and overturning. Individual grains have down-slope speeds that exceed the front speed. By contrast, avalanches of spherical glass beads (SGB) 
are quantitatively slower and smaller,
and the particles always travel slower than the front speed. We show that this difference in behavior
arises from the differing rheologies of the different particle types in
steady-state flow, and is linked to the stability of such flows. A theory based on the
nonlinear properties of the depth-averaged equations makes quantitative predictions that
agree with the experimental results and suggests that our different avalanche
structures are related to the difference between roll waves and flood waves in
hydrodynamic contexts \cite{wh1999}.  In particular, the propagation of the RNSG avalanches involves a shock, while the SGB avalanches have a continuous structure.

Two key lessons emerge from this study.  The first is that our understanding of the rheology of dense granular flows is now sufficiently robust to allow successful modeling of quite detailed dynamical phenomena such as avalanche profiles.  The second is that the quantitative differences between the rheology of the most commonly studied spherical particles and the more realistic rough particles result in  qualitative, indeed, dramatic differences in their dynamical behavior.  

A thin layer of granular material inclined at an angle $\theta$  
is stable for a wide range of $\theta$ and thickness $h$.
The onset of flow is expected only above a critical layer
thickness \cite{po1999,pore1996} $h_c$, and the flow subsides
at $h_s < h_c$.
The values of $h_s(\theta)$ and $h_c(\theta)$ decrease rapidly  
with increasing $\theta$. Thus, the layer can become unstable  
by increasing $\theta$ by a small amount $\delta\theta$   
\cite{dado1999,da2001}
or by increasing $h$ by adding new grains to the layer at a low flow  
rate, as we do here. In the former case, where the  
whole layer becomes metastable \cite{dado1999,da2001}, the shape  
and propagation of the avalanches depends critically on $\delta\theta$
such that either strictly downward or simultaneous downward  
and upward expanding avalanches can occur depending on the  
magnitude of $\delta\theta$.  
In our case a homogeneous static layer is prepared that is stable to small 
perturbations.
New grains are only added at the top region ($5\%$) of the plane in the form 
of a very low incoming flux (shower) perpendicular to the plane. 
Due to the low incoming flux 
the height slowly increases locally, and when the little pile formed in this manner becomes 
unstable an avalanche is created, which travels down the rest of the plane 
on top of the stable static layer.

The experiments reported here used an inclined plane 2.2 m long and  
0.4 m wide with a rough surface of particles glued to a glass  
substrate or sandpaper with different grit sizes. The grains were of  
two types: RNSG of sand or salt with  
several mean diameters $d$  (sand: 0.4$\pm$0.1 mm; sand: 0.2$\pm$0.1 mm;  
salt: 0.4$\pm$0.1 mm) and SGB with $d$ = 0.5$\pm$0.1 mm. The mass flow rate per unit  
width of the channel $Q$ was adjusted so that individual well-resolved avalanches  
formed, and was 0.17 g/s-cm for RNSG and 0.05 g/s-cm for SGB.   
Varying the incoming flux or the kinetic energy of the incoming particles
gives rise to to a change in the size distribution and frequency of avalanches,
which is not the subject of the present study.
The angle $\theta$ was varied  
between $32^o$ and $41^o$ for RNSG and between $22^o$ and $26^o$ for  
SGB. The differences in $Q$ and $\theta$ for the different materials  
reflect the material variations in the critical angle $\theta_c$  
and the angle of repose $\theta_r$ where flow stops. The experimental  
methods for obtaining the data presented below include:  
visualization of grain motion with high speed (1000fps) video imaging,  
determination of the lateral sizes of avalanches by image  
differencing, reconstruction of the 2D height profiles of avalanches  
using a laser sheet, and determination of particle and front velocities  
by the analysis of single particle trajectories on space-time plots.

\begin{figure}[ht]
\resizebox{70mm}{!}{
\includegraphics{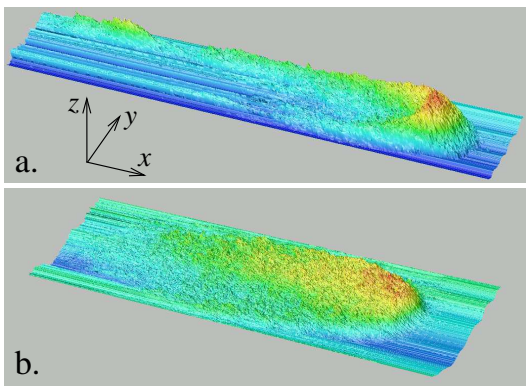}
}
\caption{
Height profiles of a) RNSG avalanche for $\theta=36.8^o$.   
Image size $7.2$   cm  x  $56$ cm, (vertical size rescaled by 25x)  
maximum height: $h_m=0.34$  cm, static layer thickness $h_s=0.12$  cm;  
b) SGB avalanche for $\theta=24.3^o$.
Image size $12.2$  cm  x  $46.8$  cm, (vertical size rescaled by 25x) maximum
height: $h_m=0.29$   cm, static layer thickness $h_s=0.18$  cm.
}
\label{avalanche3D}
\end{figure}

\begin{figure}[ht]
\resizebox{70mm}{!}{
\includegraphics{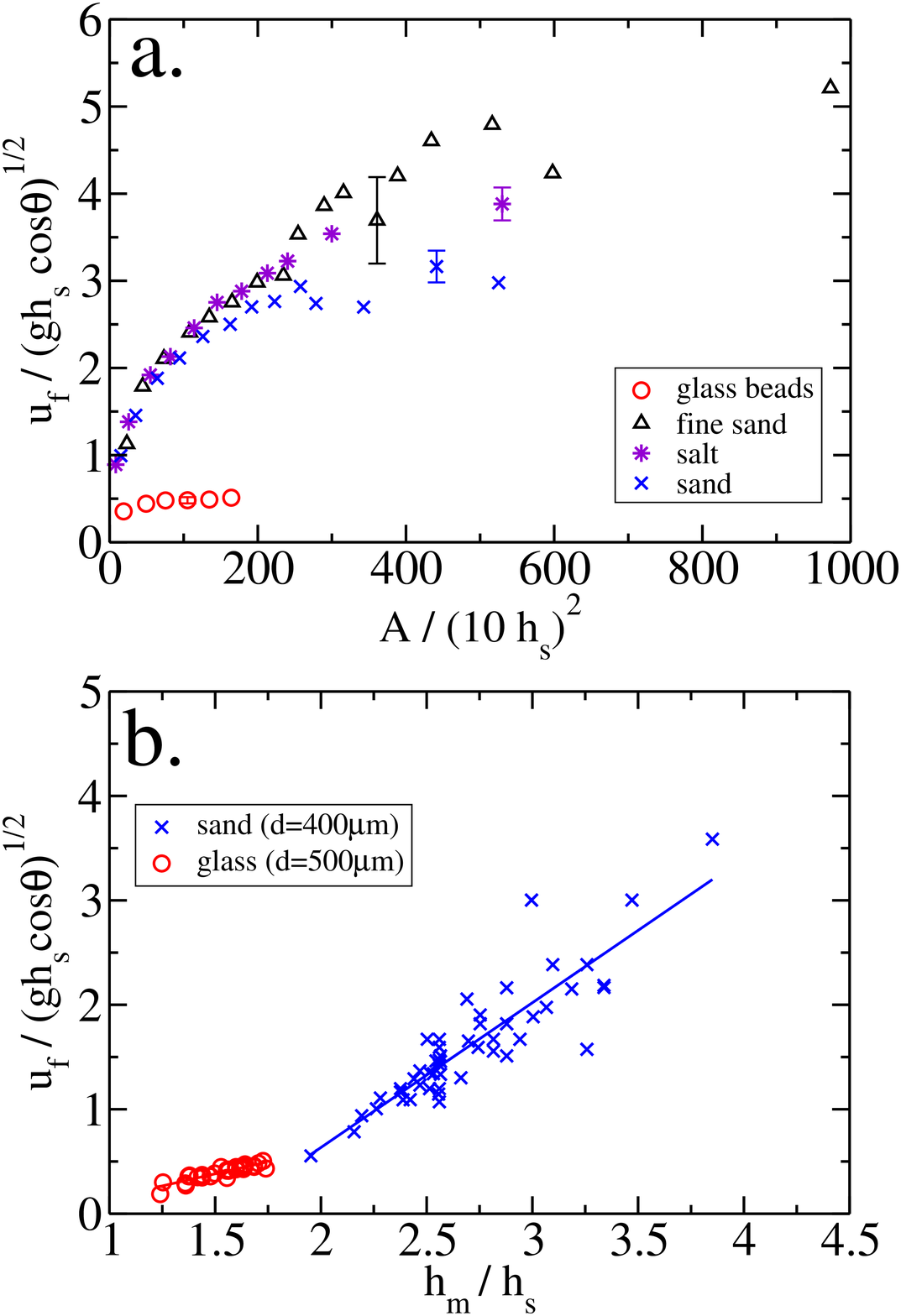}
}
\caption{Dimensionless avalanche front velocity  vs a) dimensionless  
area A and b) dimensionless maximum height $h_m/h_s$ for sand (x), glass  
beads (o), salt (*) and fine sand ($\triangle$).}
\label{size-velo}
\end{figure}

Figure \ref{avalanche3D} shows the reconstructed  
2D height profiles for sand and glass bead avalanches for operating  
conditions ($\theta$ = 36.8$^o$, 24.3$^o$ and $Q$ = 0.17 g/s-cm,  
0.05g/s-cm for RNSG and SGB, respectively) that produce roughly  
the same number of spatially-localized avalanches in an instantaneous  
image. For gravity-driven flows on an incline \cite{erha2002} the  
characteristic length scale is the height of the layer $h_s$ and  
the corresponding velocity scale is $\sqrt{gh_s \cos\theta}$.  
This normalization collapses data for different sized RNSG as  
shown in Fig.~\ref{size-velo} where the spread is about 25$\%$  
around the mean of the three data sets. For the SGB the velocities  
of avalanches with the same dimensionless areas are smaller by  
roughly a factor of 4. Further, the maximum dimensionless area  
of SGB avalanches is less than for RNSG avalanches, again by about  
a factor of 4. The corresponding dimensionless avalanche velocity  
as a function of avalanche height, normalized by $h_s$, is shown  
in Fig.~\ref{size-velo}b. The data support the hypothesis  
that there are two classes of avalanches: SGB avalanches have  
maximum height $h_m$ that is always less than 2$h_s$ whereas the height  
of RNSG avalanches is always greater than 2$h_s$.  
Our measurement of the height of avalanches relative to $h_s$  
for SGB $h_m/h_s \approx 1.45 \pm 0.1$ is very similar to  
previous measurements for granular waves \cite{da2001}  
for which $h_m/h_s \approx 1.55 \pm 0.1$, whereas the RNSG heights  
reported here are considerably higher: $h_m/h_s \approx 2.5 \pm 0.2$.

\begin{figure}[ht]
\resizebox{70mm}{!}{
\includegraphics{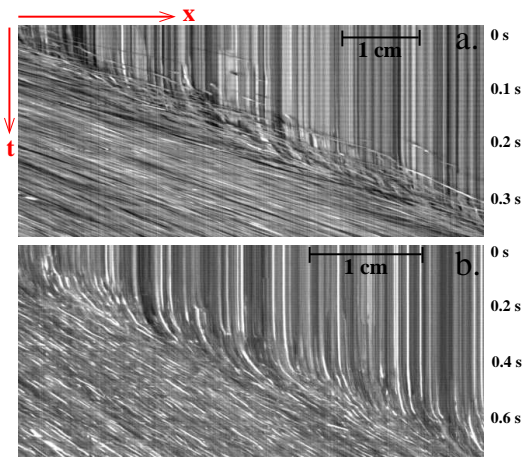}
}
\caption{Space-time plots taken along the symmetry axis of the avalanches
a) sand for $\theta=36.8^o$ with scaled velocities (Froude numbers)  
$\Froude_f=2.12$, $\Froude_p=2.55$, and   
b) glass beads $\theta=23.3^{o}$ with $\Froude_f=0.55$,  $\Froude_p=0.33$.
}
\label{spatio}
\end{figure}

\begin{figure}[ht]
\resizebox{70mm}{!}{
\includegraphics{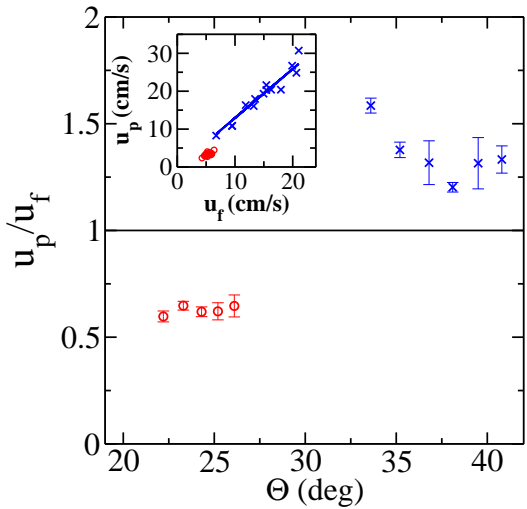}
}
\caption{ Ratio of particle and front velocities $u_p/u_f$ for glass  
beads (o) and sand (x). The inset shows $u_p$ vs $u_f$ for $\theta=36.8^o$
(sand) and   $\theta=25.2^o$ (glass beads).
}
\label{velofront-particle}
\end{figure}

The origin of this dramatic difference is revealed in detailed  
measurements of the velocity distribution of surface particles  
for the two types of avalanches.  The avalanche front speed $u_f$  
and the surface mean particle speed $u_p$ right behind the front  
can be determined by measuring the image intensity along the  
centerline of an avalanche. Plotting the intensity in a space-time  
plot, see Fig.~\ref{spatio}, reveals streaks associated with
particles and a well-differentiated front for each avalanche  
type. For RNSG avalanches (Fig.~\ref{spatio}a) the streaks  
within the avalanche are less steep than the front line, indicating  
$u_p > u_f$, whereas the opposite is true (Fig.~\ref{spatio}b)  
for the SGB avalanches. In Fig.~\ref{velofront-particle}, the  
plot of $u_p$ versus $u_f$ shows that for SGB $u_p < u_f$, with a  
ratio $u_p/u_f \approx 0.7$, whereas for RNSG avalanches  
$u_p > u_f$ with $u_p/u_f  \approx 1.4$. Another feature of  
the avalanches seen in Fig.~\ref{spatio} is the  
continuous form of SGB avalanches as evidenced by the curved  
paths just ahead of the front (representing the acceleration of  
particles in this region) as compared with particles being thrown  
out of the main body of the avalanche for the RNSG type.  High speed  
imaging demonstrates \cite{movie} that an RNSG avalanche consists of a fast  
moving packet of grains that overtakes the front, like a breaking  
wave in a fluid, which entrains granular material from the  
underlying layer as it passes over.  SGB avalanches, on the other  
hand, are continuous in that the static stresses that  
hold grains in place in front of the   
granular packet change and the previously stable 
packing  collapses. 

These two distinct types of avalanches can be understood as  
consequences of the underlying rheology of the RNSG and SGB systems  
and of the structure of the dynamical equations for a propagating  
front. Let us first consider the rheology.  
For steady-state flows, both sand and glass beads have depth-averaged  
velocities $u$ given by the Pouliquen form  
\cite{po1999,sier2001,erha2002}

\begin{equation}
\frac{u}{\sqrt{gh}} = F(h/h_s(\theta)), \ \ {\rm where} \ \ F(y)=\beta y -\gamma,
\label{Pouli1}
\end{equation}

\noindent where $h_s(\theta)$ is the minimum height of a flowing pile for an angle  
$\theta$. (Note that the depth-averaged quantities of the continuum  
model are different than the surface velocities measured in the  
experiments.) For sand \cite{fopo2003}, the coefficients in the function $F(y)$ are  
$\beta \approx \gamma \approx 0.7$, whereas for glass beads,  
$\beta \approx 0.14$ and  
$\gamma \approx 0$ \cite{fopo2003,po1999,sier2001}.  
We now apply these results to our RNSG and SGB systems,  
respectively.

The simplest dynamical picture of granular flows down an incline  
is achieved using the Saint-Venant shallow flow equations, adapted  
for granular media by Savage and Hutter \cite{sahu1989}. For a flow  
of height $h$ and mean velocity $u$, these describe flow down a  
plane, with the plane parallel to the $x$-direction, by the  
averaged equations

\begin{equation}
\frac{\partial h}{\partial t} + \frac{\partial (hu)}{\partial x}=0;
\label{saintvenant1}
\end{equation}

\begin{equation}
\frac{\partial (hu)}{\partial t}+  
\alpha \frac{\partial (hu^2)}{\partial x} =
\left( \tan \theta -\mu (u,h) -K \frac{\partial h}{\partial x} \right) gh \cos \theta.
\label{saintvenant2}
\end{equation}

\noindent Here $\alpha$ is determined by the profile of the flow,  
$\alpha=1$ for plug flow (as in Ref. \cite{sahu1989}),
$\alpha = 4/3$ for a linear flow profile, or 5/4 for a convex  
Bagnold profile \cite{doan1999}. The parameter $K$ is determined by the ratio of the
normal stresses in the flow: the stress parallel to the bed, $\sigma_{xx}$, 
and that perpendicular to the bed, $\sigma_{zz}$.  
Numerical results show that $K \equiv \sigma_{xx}/\sigma_{zz} \approx 1$ 
for steady-state flows \cite{sier2001}. 
The friction coefficient $\mu(u,h)$ is determined  
by the requirement that the steady flow obey the rheology shown  
in Eq.~(\ref{Pouli1}), and will thus vary with the particle type.

Since the dimensionless velocity, typically given as the  
Froude number ${\Froude} = u/\sqrt{gh\cos\theta}$, is small near the  
critical angle or angle of repose for both the sand and the  
glass bead flows, it is natural to take the limit $\Froude  \rightarrow 0$ in Eqs.~(\ref{saintvenant1}, \ref{saintvenant2}),
which suppresses  
the LHS of Eq.~(\ref{saintvenant2}).  
We thereby obtain with some work an approximate equation for $h$:

\begin{equation}
\frac{\partial h}{\partial t}+
a(h) \frac{\partial h}{\partial x} =  
\nu \frac{\partial^2 h}{\partial x^2}; \ \  
a(h) = \sqrt{gh}  
\left(\frac{5}{2} \beta \frac{h}{h_s(\theta)} -\frac{3}{2}\gamma\right),
\label{resulth}
\end{equation}

\noindent where the ``viscosity" $\nu$ can be computed as $\nu\sim d\sqrt{gh}$  
\cite{OPprivate}. This equation has solutions similar to those  
of Burger's equation, with a Burger's shock smoothed  
by the influence of the viscosity term. Thus, there is a solution  
consisting of a single hump propagating down the slope with velocity  
$a(h)$, with a smooth structure determined by the competition  
between this nonlinear velocity term on the LHS of Eq.~(\ref{resulth})
and the viscosity term.

Turning to the full system of  
Eqs.~(\ref{saintvenant1}-\ref{saintvenant2}), however, we observe  
a potential flaw in this approach \cite{wh1999}. This full system  
is hyperbolic with characteristic velocities

\begin{equation}
c_{\pm}=u\left(\alpha \pm \sqrt{\alpha (\alpha-1) +\frac{K}{\Froude^2}}\right).
\label{resultc}
\end{equation}

\noindent If the velocity appearing in Eq.~(\ref{resulth}) does not  
obey  $a < c_+$, then Eq.~(\ref{resulth}) predicts a structure that  
moves faster than the maximum rate at which information can be  
propagated in the full system of equations, which is clearly  
impossible. In these circumstances, the Burger's type solution  
transforms itself into a truly discontinuous solution traveling at  
velocity $c_+$ \cite{wh1999}, which is described by the full system Eqs.~(2-3) rather than by Eq.~(\ref{resulth}).

Using the rheology determined by Eqs.~(\ref{Pouli1}),  
one can see that for SGB, taking a value of $\alpha=4/3$, this  
discontinuous solution will only develop for heights above  
$h/h_s\approx 6$, whereas for RNSG, it will develop as soon as  
the avalanche has a height $h/h_s\approx 1.3$ (the precise  
threshold depending on the value of $\theta$). This quantitative  
height condition is consistent with the data in Fig.~\ref{size-velo}b  
where $h_m/h_s < 6$ for SGB and $h_m/h_s >1.3$ for RNSG. Further,  
calculation of the experimental ratio $a/c_+$ using  
the rheology of Eqs.~(\ref{Pouli1}) yields  
$a/c_+ < 0.75$ for SGB and $a/c_+ > 1.1$ for RNSG,  
which is also consistent with this picture. Thus, we conclude  
that the glass bead avalanches reflect smooth solutions of  
Eq.~(\ref{resulth}), with $a < c_+$, whereas the sand avalanches  
represent discontinuous solutions of the full system,  
traveling at velocity $c_+$. The latter avalanches propagate  
into a quiescent bed because they are traveling at the  
characteristic velocity for the medium.  
The glass bead avalanches are analogous to ``flood waves" in river flows,  
whereas the sand avalanches are analogous to ``roll waves" in these  
flows \cite{wh1999,dr1949}.
Note that ahead of the flowing avalanche, the moving material propagates 
into a material at rest, which is presumably in a state close to the 
critical Mohr-Coulomb state \cite{nedderman}. Unlike the flowing state, 
for which $\sigma_{xx} \approx \sigma_{zz}$, in this critical state 
$\sigma_{xx} > \sigma_{zz}$. Thus the transition region, in which 
the flow accelerates from rest into a pseudo-steady state described 
by the continuum theory, can be viewed as a region of passive Rankine 
failure, through which the compressive stress parallel to the bed, 
$\sigma_{xx}$, is decreasing with time. The mechanics of this region 
is complex, and cannot be described by the Saint-Venant equations alone.

Finally, we point out that in the linear theory of the instability of  
steady flows, developed by Forterre and Pouliquen, the criterion  
$a < c_+$ corresponds to the stable regime of these flows with  
respect to wave disturbances \cite{fopo2003}. Thus our observation  
of discontinuous avalanches for sand and smooth avalanches for  
glass beads dovetails nicely with their observation that steady  
flows of sand are far more unstable to such disturbances than are  
steady flows of glass beads.

We thank Brent Daniel, Michael Rivera, Ell\'ak Somfai, and Ben White
for helpful scientific conversations.   
Experimental work was funded by the US Department of  
Energy (W-7405-ENG-36).

\end{document}